\documentclass[]{aa} 

\usepackage{graphicx}
\usepackage{times}

\begin{document}
   \title{Isotopic abundance of nitrogen and carbon in distant comets%
\thanks{Based on observations collected at the European 
Southern Observatory, Paranal, Chile 
(ESO Programmes 270.C-5043, 073.C-0525 and 274.C-5015).}}


   \author{J. Manfroid
          \inst{1}\fnmsep\thanks{Research Director FNRS}
          \and
E. Jehin
\inst{2}\and
D. Hutsem\'ekers 
\inst{1}\fnmsep\thanks{Research Associate FNRS}\and
A. Cochran 
\inst{3}\and
J.-M. Zucconi  
\inst{4}\and
C. Arpigny     
 \inst{1}\and
R. Schulz  
 \inst{5}\and
J.A. St\"uwe  
\inst{6}
}

   \offprints{J. Manfroid}

\institute{Institut d'Astrophysique et de G\'eophysique,
Universit\'e de Li\`ege, All\'ee du 6 ao\^ut 17, B-4000 Li\`ege
\and
European Southern Observatory, Casilla 19001, Santiago, 
Chile
\and
Department of Astronomy and McDonald Observatory, University 
of Texas at Austin, C-1400, Austin, USA
\and
Observatoire de Besan\c{c}on, F25010 Besan\c{c}on Cedex, France
\and
ESA/RSSD, ESTEC, P.O. Box 299, NL-2200 AG Noordwijk, 
The Netherlands
\and
Leiden Observatory, NL-2300 RA Leiden, The Netherlands
}

   \date{}

   \abstract{
The  $^{12}$C$^{14}$N/$^{12}$C$^{15}$N\ and $^{12}$C$^{14}$N/$^{13}$C$^{14}$N\ 
isotopic ratios have been determined
in comets C/1995 O1 (Hale-Bopp), C/2001 Q4 (NEAT) and
C/2003 K4 (LINEAR)      at  heliocentric distances of, respectively,
2.7, 3.7 and 2.6 AU.  These ratios have also been measured at $r\sim1$ AU.
No significant differences were found 
between all determinations, nor  with the value obtained for
other comets. If confirmed, 
the discrepancy between the nitrogen isotopic ratios 
from optical and millimeter
measurements on CN and HCN  would
rule out HCN as a major parent
of the cometary CN radicals.

   }

   \maketitle
%

\section{Introduction}

Determination of the abundance ratios of the stable isotopes of the
light elements in different objects of the solar system provides
important clues in the study of its origin and early history. Comets
carry most valuable information regarding the material in the primitive
solar nebula.

   \begin{figure}
\includegraphics[width=88mm]{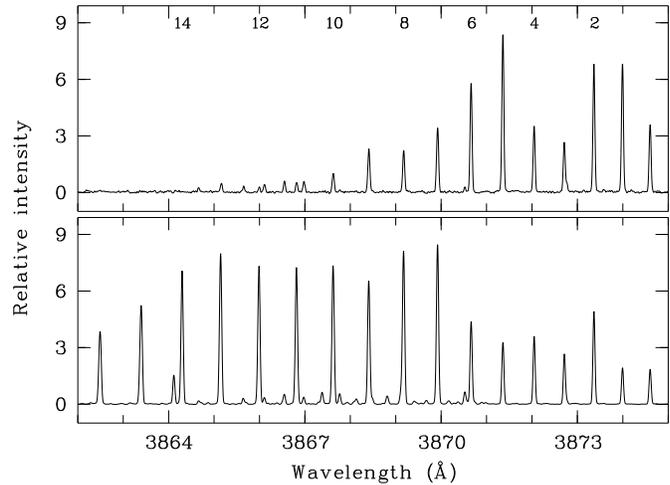}
\caption{Spectra of comet C/2001 Q4 at 3.7 AU (top) and 0.98 AU
showing part of the R branch of the CN B-X (0,0) band.
The corresponding quantum numbers are indicated at the top.
The major differences are related to the heliocentric distance and show up in
the excitation of lines of higher quantum number when closer to the Sun.
The intensity scale is relative only.              }
         \label{Figa}
   \end{figure}

We identified for the first time a number
of emission features belonging to the $^{12}$C$^{15}$N B-X (0,0) band
in
spectra obtained with the SOFIN
spectrograph at the Nordic Optical Telescope (Canary Islands). This allowed
us to make the first optical measurement of the nitrogen isotopic ratio
$^{14}$N/$^{15}$N\ in a comet (Arpigny et al.~\cite{Arpigny}). This ratio was
found to be lower by a factor of about two than the terrestrial
value (272) and,
less than half those obtained in Hale-Bopp from millimiter
measurements of HCN, a possible main parent of CN
($^{14}$N/$^{15}$N$=323\pm46$ at $r=1.2$~AU: Jewitt et al.~\cite{Jewitt}, 
$330\pm98$, $r=0.92$~AU: Ziurys et al.~\cite{Ziurys}). 
Spectra of fainter comets obtained with the ESO VLT 
and the 2.7m McDonald telescopes gave   similar results, viz
$140\pm25$ and $170\pm50$ for comets 122P/de Vico ($r=0.66$~AU) and 153P/Ikeya-Zhang
($r=0.92$~AU) respectively
(Jehin et al.~\cite{Jehin}).
On the other hand the optical determinations of the $^{12}$C/$^{13}$C\ ratio
consistently yield values around 90, only slightly lower than  the
HCN millimiter measurements (respectively $111\pm12$ and $109\pm22$ from
the above-cited authors), and in agreement with the solar value (89).    

The differing results for nitrogen in CN and HCN would indicate   that
cometary CN radicals are produced
from at least one other source with a much lower
N isotopic ratio.

A monitoring of Hale-Bopp showed        that
somewhere beyond about 3 AU, the production rates of HCN and CN agree
with one another within certain limits (Rauer et al.~\cite{Rauer}).
Hence,          sufficiently far away from the Sun, HCN could be the
only source of CN such that $^{14}$N/$^{15}$N from CN should match   
the HCN value.  
In order to investigate this possibility, we measured the N
isotopic ratio in three comets at such heliocentric distances.
By comparison, our previously published data concern comets
much closer to the Sun (0.6--1.3 AU).

\section{Observations and data analysis}            

Comets
C/1995 O1 (Hale-Bopp), C/2001 Q4 (NEAT) and 
C/2003 K4 (LINEAR) have been observed at heliocentric distances 
of, respectively,
2.7, 3.7 and 2.6 AU, and all three  were also observed close to
perihelion.
Various telescopes were used in Chile, the USA, and the Canary Islands.
Spectra of the CN violet band were obtained at   
resolving power of 60000 or higher
(see Table~\ref{observ}).

   \begin{table*}
      \caption[]{Observational data. 
$r$ is the heliocentric distance.
$m_{\rm r}$ is the heliocentric visual magnitude.
2dc is the McDonald 2D-coude spectrograph, SOFIN
and UVES are 
the high-resolution spectrographs at, respectively, the NOT and       ESO VLT telescopes.
$n$ is the number of spectra obtained, for 
a total exposure time of Exp. Extent is
the size of the zone explored 
at the distance of the comet.
$R$ is the instrumental resolving power. Fwhm
is the measured width of the CN lines. 
}
         \label{observ}
\includegraphics[width=18cm, viewport=40 510 520 670, clip]{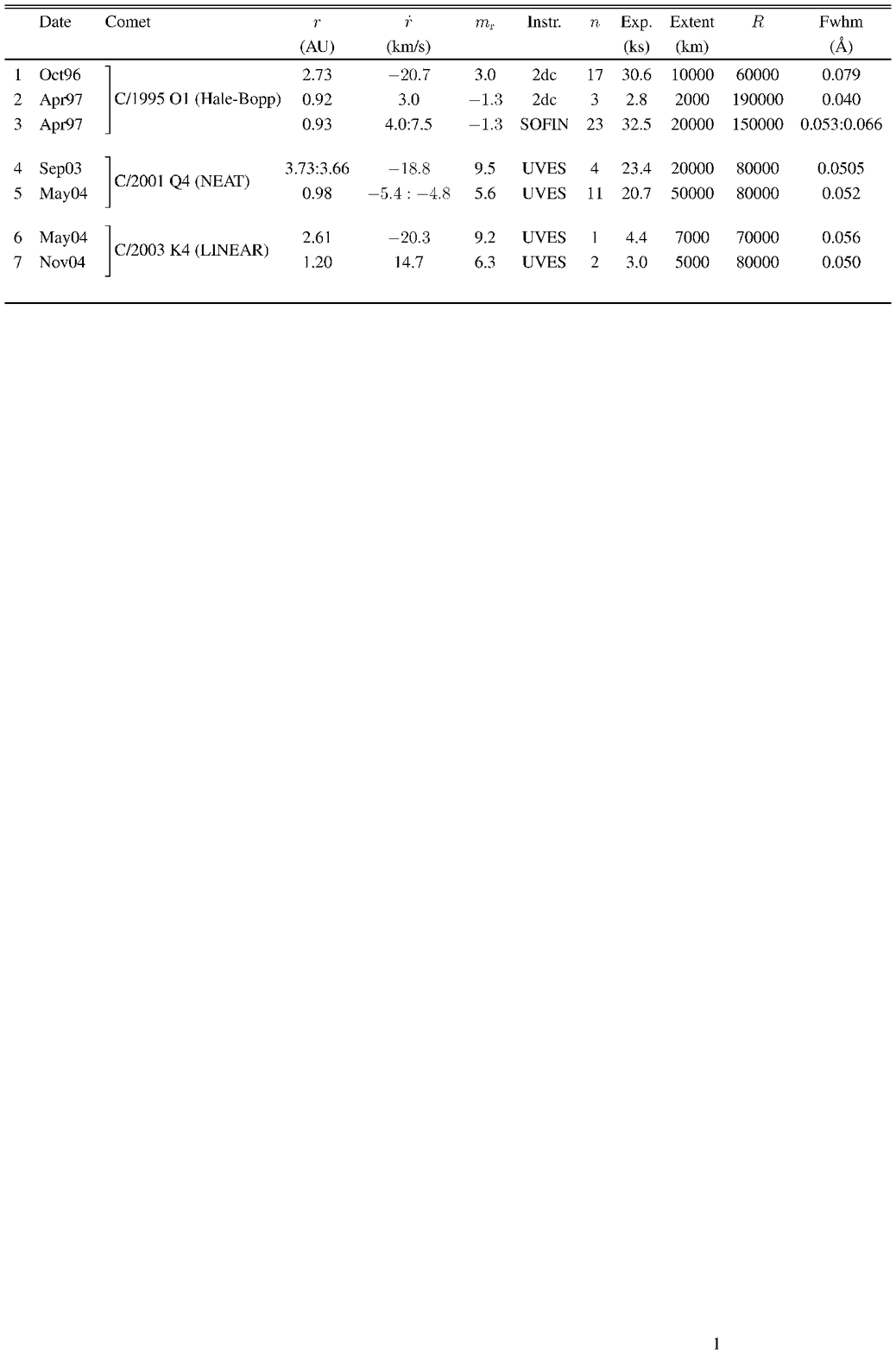}
   \end{table*}

The techniques used to reduce the data and 
measure the isotopic ratios are described
in Arpigny et al.~(\cite{Arpigny}) and 
Jehin et al.~(\cite{Jehin}). The $^{12}$C/$^{13}$C\ and $^{14}$N/$^{15}$N\ ratios are obtained 
simultaneously  by fitting the spectra with theoretical models.
An empirical correction was adopted to take account of additional effects not
covered by the pure fluorescence model (collisional and possibly other
unidentified effects).

Indeed, the comparison of theoretical and observational $^{12}$C$^{14}$N{} data for
several comets shows that a 3-parameter function, with the quantum number,
$N$, as the independant variable, always brings a satisfactory agreement. The
intensity $f_{N}$  of line $N$ is corrected in the following way:
\begin{equation}
F_{N} = f_{N} ~ e^{cN} / (e^{(N-a)/b}+1)
\label{eq0}
\end{equation}

{\noindent with $a,b,c$ being estimated through an iterative process, for each
individual spectrum.
The same correction is applied to $^{13}$C$^{14}$N{} and $^{12}$C$^{15}$N{} under the assumption that
these spectra are similarly affected.}
These corrections typically stay within 15\%. In cases of severe collision effects,
in the central region of the coma of C/2001 Q4, they can reach up to 40\%.
One spectrum taken on May 6, 2004, right on the nucleus shows corrections of
from $-65$\% at R(0) up to $+65$\% around R(15).
The rms deviation of the fit between the observations and the corrected
model is generally between 5 and 15\%, i.e., 3 to 4 times smaller
than for the uncorrected model.

Only the R lines of the B-X (0,0) band (i.e., shortward of 3875 \AA)
are  used since  the P lines of the three isotopes are strongly
blended.
The spectra obtained at large heliocentric distance have relatively
low S/N ratios. This makes   the use of a global fitting technique
mandatory.  As seen in Fig~\ref{Figa} the bulk of the CN emission is
concentrated in fewer lines, namely those  of low quantum number, when 
farther from the Sun.
This increases the overall S/N ratio for the isotopic lines since
the noise is dominated by the background and the CCD readout.
Simultaneously, the low-excitation spectra are less contaminated
by the P lines of the (1,1) band of the dominant species.

   \begin{figure}
   \centering
   \includegraphics[width=8cm, trim=-0 0 0 0, clip]{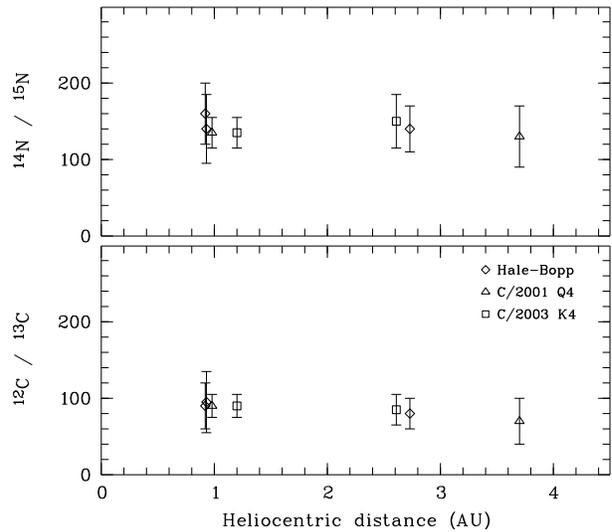}
      \caption{Nitrogen and carbon isotopic ratios at various heliocentric distances}
         \label{Figd}
   \end{figure}

The estimated isotopic ratios are presented 
in Table~\ref{result} and plotted versus the heliocentric distance
in Fig~\ref{Figd}. The values based on    the Hale-Bopp
SOFIN observations are slightly different from (and supersede) those given 
by Arpigny et al.~(\cite{Arpigny}). This is mainly due to 
the use of better molecular parameters in the fluorescence model      
and of the ad-hoc empirical correction.
The Hale-Bopp SOFIN values agree with the ratios
 from McDonald spectra obtained
quasi simultaneously. 

\section{Discussion}
Our measurements show that changes in the $^{14}$N/$^{15}$N\ and $^{12}$C/$^{13}$C\ ratios 
with heliocentric distance are small as 
determined from the CN radical. This means that the CN isotopic
mixture remains constant within the explored range 
of heliocentric distances, and
is the same for every comet, with average values of, respectively,
86 and 141   for $^{12}$C/$^{13}$C\ and $^{14}$N/$^{15}$N. 
The carbon ratio is compatible with previous measurements, including
millimiter measurements in HCN. On the other hand the nitrogen
 ratio is compatible with
optical measurements of CN in other comets, but not with 
the HCN observations of Hale-Bopp.

In order to get some idea of the constraints on the actual isotopic
variations, we assume two sources of CN, namely HCN and X.
Let us call
$ \zeta = [^{12}{\rm C}^{14}{\rm N}/^{12}{\rm C}^{15}{\rm N}]_{\rm total}$  
the observed isotopic ratio of CN, and   
$ \zeta_{\rm H}= [^{12}{\rm C}^{14}{\rm N}/^{12}{\rm C}^{15}{\rm N}]_{\rm HCN}$    
the isotopic ratio of the CN radicals coming from HCN.
Similarly, we define
$ \zeta_{\rm X} = [^{12}{\rm C}^{14}{\rm N}/^{12}{\rm C}^{15}{\rm N}]_{\rm X}$.  

We call
$\epsilon =\, $CN$_{\rm HCN}/$CN$_{\rm total}
 \approx \,^{12}{\rm C}^{14}{\rm N}_{\rm HCN}/^{12}{\rm C}^{14}{\rm N}_{\rm total}$
the fraction of CN coming from HCN.

\medskip

One has 
\begin{equation}          
    1/ \zeta\approx\epsilon (1/\zeta_{\rm H}-1/\zeta_{\rm X}) + 1/\zeta_{\rm X}\,, \label{eqa}
   \end{equation}
so that
\begin{equation}
(1/\zeta_1-1/\zeta_2)\approx(\epsilon_1-\epsilon_2)(1/\zeta_{\rm H}-1/\zeta_{\rm X})\,, \label{eqb}
 \end{equation}
where subscripts 1 and 2 refer to different observations --- e.g.,
the same comet at different heliocentric or nucleocentric distances, or 
different comets. Hence, assuming that $\zeta_{\rm X}$ and $\zeta_{\rm H}$ do not vary, 
changes in the observed ratio $1/\zeta= \,^{12}{\mathrm{C}}^{15}{\rm N}/^{12}{\rm C}^{14}{\rm N}$ 
are directly proportional
to $\epsilon_1-\epsilon_2$, i.e., to changes in the relative importance of the 
column density of CN$_{\rm HCN}$ or, with the opposite sign, of CN$_{\rm X}$.
 
Let us consider an  isotopic mixture 
with the observed ratio  $\zeta_1=140$ and an  isotopic ratio 
$\zeta_{\rm H}=325$ corresponding to the millimiter observations of HCN.
Equation~(\ref{eqa}) links   $\zeta_{\rm X}$ and $\epsilon_1$.
Equation~(\ref{eqb}) can be used to calculate the 
proportion $\epsilon_2$ needed to produce a significant change in the
observed ratio, say $\zeta_2=200$, which would be easily detected.

First, we assume that $\zeta_1=140$ characterizes a 
mixture of CN$_{\rm HCN}$ and CN$_{\rm X}$ in equal 
proportion ($\epsilon_1=0.5$). The isotopic ratio of X is then  $\zeta_{\rm X}=89$.
The proportion of CN$_{\rm HCN}$ must increase to $\epsilon_2=0.76$
in order to yield an overall $\zeta_2=200$.

As another (extreme) example, let us assume that the  observed overall 
$\zeta_1=140$     characterizes CN coming exclusively from X ($\epsilon_1=0$). 
The $^{14}$N/$^{15}$N\ ratio  $\zeta_2=200$ is reached when 53\% of CN
comes from HCN ($\epsilon_2=0.53$).

Our observations show that there is no such change in composition up to 3.7 AU.

Then, the hypothetical transition zone between a  pure HCN source and a 
multiple source of CN must  be farther out than 3.7 AU.
This does not fit well with Rauer et al.'s~(\cite{Rauer}) results. 

From Eq.~(\ref{eqb}), imposing  
the constancy of  the isotopic ratios measured
in several comets  at various distances  ($\zeta_1=\zeta_2$), we find: 
\newline
(i) 
$\zeta_{\rm X}=\zeta_{\rm H}=\zeta\,(=140)$,
isotopically homogeneous sources, no constraints on
the mixture proportions, 
or 
\newline
(ii)
$\epsilon_1=\epsilon_2$,
the mixtures have identical proportions,

We shall not discuss further hypothesis (ii) because
a highly improbable conjunction of coincidences would be
needed to achieve the constancy of the fractions $\epsilon$
in every comet, under any conditions.

The first hypothesis (i) implies that either the observed
HCN or CN  nitrogen isotopic ratio is faulty, or that the observed CN
does not come from HCN. 

In the optical domain, the measurement of the CN isotopic 
ratios is straightforward 
as soon as the S/N ratio is high enough. The number of lines
involved (a minimum of 7 in the present analysis) makes the 
possible influence of underlying
unidentified features rather improbable. The fluorescence
models appear to be reliable, so errors by a factor 2 seem
to be excluded. 

The HCN isotopic ratio is difficult to measure, and 
so far has been obtained in only one comet, the exceptionally
bright Hale-Bopp, but the two independent measurements
agree with one another and there is          no reason to
discard them. If there was a problem, the carbon ratio
should probably show it in similar proportions. Nevertheless, it 
would be important to see whether the
HCN results can be confirmed on a broad sample of
comets in order to pin-point the nature of the problem of
the isotopic ratio of N. Are there actually large 
N isotopic differences between the various constituents of a
comet? Or does the cometary material have an overall
excess of $^{15}$N\ as compared to our current estimates 
of the solar nebula? Did the problem arise in this
primitive nebula or is it linked to the formation and history
of the cometary bodies?

The discrepancy between the observed CN and HCN isotopic ratios
could be explained if most of the CN released by HCN is produced   
outside of the zones explored by the
observations, i.e., farther than about $5\,10^4$ km from  the nucleus.
HCN would then photodissociate slowly, much farther than expected
from the nucleus. Its contribution 
to CN along the line of sight toward more central regions
of the coma would be  negligible. 
A sufficiently low dissociation rate would
make the corresponding CN invisible even at large
heliocentric distances. 
However, the spatial distribution of HCN is rather steep and 
excludes such large scale lengths (see, e.g. Irvine et al.~\cite{Irvine}). 
The major sources of CN --- at small and large heliocentric distances --- must then be
some other molecule(s) with a dissociation scale length rather  
comparable to that of HCN and a lower $^{14}$N/$^{15}$N.
 
Indeed, other sources of CN --- simple or complex organic molecules --- 
have been mentioned long ago in the literature, notably  C$_2$N$_2$ (see, e.g., 
Swings \&  Haser \cite{Swings}). 

The difficulties met when investigating 
HCN and CH$_3$CN as possible parents of CN were raised  by
Bockel\'ee-Morvan \& Crovisier (\cite{Bockelee}) on the basis
of the expansion velocity in the coma.    
They suggested C$_2$N$_2$ and HC$_3$N as possible alternatives.
After re-examination of many
near-UV observations of the CN radicals and
high-dispersion spectra of the CN Violet (0-0) band 
in the coma of comet C/Austin 1989 X1,
Festou et al.~\cite{Festou} favor a single parent which
is neither HCN nor  CH$_3$CN, but 
a molecule having a lifetime of about $3.5\, 10^4$~s 
at 1 AU. They propose the chemically stable C$_2$N$_2$ 
as a likely candidate.
Following the same lead, Bonev and Komitov (\cite{Bonev})
find that  C$_2$N$_2$ is indeed to be preferred over HCN as a single
parent of CN.
A recent analysis of the CN profile in comet 21P/Giacobini-Zinner by Lara et al.\
(\cite{Lara})
showed that HCN should be ruled out as the sole parent of CN.
On the other hand, several analyses present HCN as the major
or sole
source of CN (e.g., Korsun \& Jockers, \cite{Korsun}; Lara et al.~\cite{Larab}).

Fray et al.~(\cite{Fray}) propose HCN polymers as a source of CN.
A'Hearn et al.~\cite{AHearn} provide strong evidence that most CN
are produced from grains in the coma rather than from nuclear ices.
The link with CHON particles was suggested.
However, the correlation between 
the dust and CN distribution in the comae is not perfect. This was
shown by A'Hearn et al.~(\cite{AHearna}) in the case of comet P/Halley. 
In a study of comet Hale-Bopp, Woodney et al.~(\cite{Woodney}) 
find that there is a better correlation between HCN and CN than 
between HCN and the optically dominant dust.
A dust source could then be 
small or/and low-albedo  undetectable particles instead of the observable dust,
like black HCN polymers (e.g. Rettig, et al.~\cite{Rettig}).  
A very dusty comet like Hale-Bopp shows the same isotopic characteristics
as quasi  dust-free comets (e.g., 122P/de Vico, see Jehin et al.~\cite{Jehin}) 
and the latter do not show abnormally low CN production rates.
Greenberg and Li (\cite{Greenberg}) conclude that the source of CN and some
other simple molecules is 
the organic component in comet dust.  

   \begin{table}
      \caption[]{The carbon and nitrogen isotopic ratios, and the corresponding
estimated errors in parentheses.}
         \label{result}
\includegraphics[width=6cm, viewport=0 440 260 640, clip]{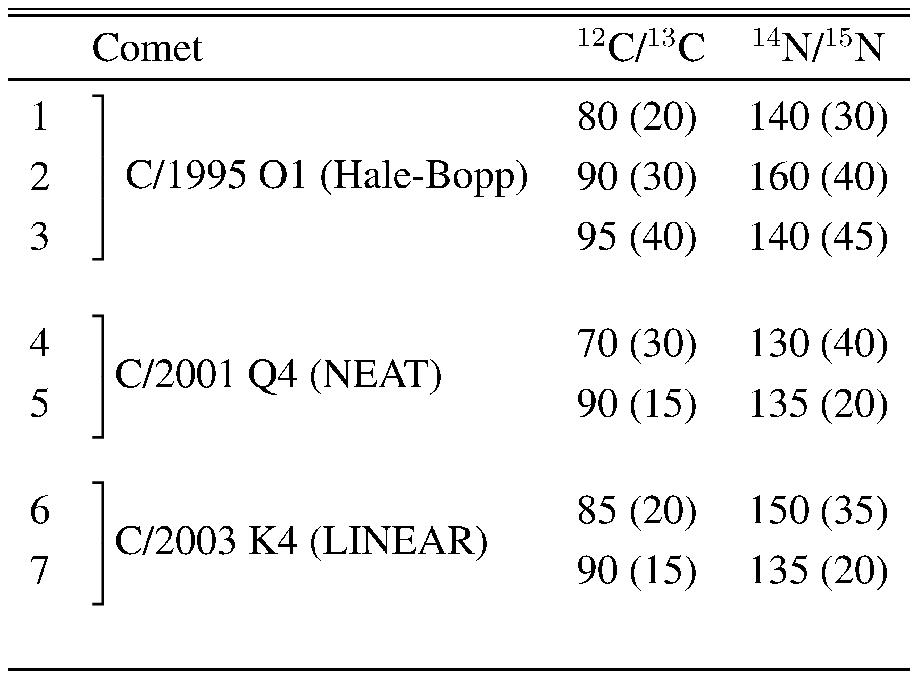}
   \end{table}

The IDPs with elevated $^{15}$N point toward the presence of organic
molecules, e.g., aromatic hydrocarbons (Keller et al.~\cite{Keller}). 
These may be related to the organic component of the cometary grains
and explain the high $^{15}$N\ abundance (see also Arpigny et al.~\cite{Arpigny}).

The link between CN and HCN in cometary comae has generally been
investigated by looking for similarities between 
scale lengths and between production rates. 
This yields different results in different comets,
a variability which would show up in the isotopic ratios.
The isotopic measurements provide an independent
and powerful fingerprinting method. If confirmed, they would show 
that HCN is not the unique --- and probably
not the major --- parent of CN at small as well as large 
heliocentric distances.  

The origin of CN is still a puzzle, as is the apparent isotopic
differentiation within cometary constituents and its consequences
on our understanding of the early solar system. Additional 
determinations of the N isotopic ratio
in HCN and  other N-bearing molecules are highly desirable
in order to shed light on these issues.

\end{document}